\documentclass[twocolumn]{jpsj3} %% for letters
\usepackage{txfonts}
\usepackage{color}

\usepackage{xspace}
\newcommand{\msr}{$\mu$SR\xspace}
\newcommand{\onlinecite}[1]{\xspace\nocite{#1}\citenum{#1}} 

\title{Toward the computational prediction of muon sites and interaction parameters}

\author{Pietro \textsc{Bonf\`a}$^{1}$\thanks{E-mail address: pietro.bonfa@fis.unipr.it}, Roberto \textsc{De Renzi}$^{1}$\thanks{E-mail address: roberto.derenzi@unipr.it}}

\inst{$^{1}$Department of Physics and Earth Sciences, University of Parma, Italy}

\abst{
The rapid developments of computational quantum chemistry methods and supercomputing facilities motivate the renewed interest in the analysis of the muon/electron interactions in $\mu$SR experiments with \emph{ab initio} approaches. Modern simulation methods seem to be able to provide the answers to the frequently asked questions of many \msr experiments: where is the muon? Is it a passive probe? What are the interaction parameters governing the muon-sample interaction?
In this review we describe some of the approaches used to provide quantitative estimations of the aforementioned quantities and we provide the reader with a short discussion on the current developments in this field.}

\kword{$\mu$SR, density functional theory, ab initio, crystal impurities, magnetism, superconductivity, hydrogen in semiconductors}

\begin{document}
\maketitle

\section{Introduction}

Muon spin rotation and relaxation spectroscopy (\msr) and density functional theory (DFT) were first theorized, and later implemented, roughly in the same years.\cite{PhysRev.136.B864,PhysRev.140.A1133,PhysRev.105.1415} Indeed, during the development of \msr as an experimental technique for studying magnetism in solid state physics, the analysis of the experimental data has greatly benefited from the theoretical insights provided by the first embryonic density functional based simulations.\cite{PhysRevLett.40.264}

To the best of our knowledge, the first example of this kind dates back in 1975.\cite{meier1975electron} Dr. P. Meier provided the first results from simulations aiming at identifying the effect of a positively charged interstitial muon in elemental metals.
From then on, many works\cite{PhysRevLett.40.264,Jena_1979} presented and discussed \emph{ab initio} methods to tackle some common sources of uncertainty that stem from complicated \msr experiments. 
We summarize them with the following three questions: {\em i)} where is the muon? {\em ii)} Can we estimate the parameters of the $\mu-e^{-}$ interaction Hamiltonian? {\em iii)} Is the muon a passive probe?

A renewed effort in trying to answer the above questions with first principles simulations has begun a few years ago.
Indeed, what has essentially changed from the 70s is our capabilities in simulating the electronic properties of complex materials, strongly reducing the impact of the approximations that must be adopted to solve the many body electronic (and in some cases also nuclear) problem on the parameters under investigation.
It is known, for example, that DFT is very accurate in determining the bond distances and unit cell sizes even when adopting the Local Density Approximation (LDA) for the exchange and correlation potential.
This is the rougher approximation, that disregards the non-local effects of the exchange interaction, and there are many cases where it is not sufficient to obtain a realistic description of the material.

Starting from the LDA, the Generalized Gradient Approximation generally improves the description of the lattice positions. From there, in recent years, many rungs have been added to the ``Jacob's ladder of density functional approximations for the exchange-correlation energy'' \cite{vxcladder}. Moreover, many body approaches different from DFT and \emph{ad hoc} corrections to the Kohn-Sham Hamiltonian have greatly improved the capabilities of the density functional method in describing the electronic degrees of freedom of crystals and molecules. The reader is referred to one of the many review articles that discusses this topic.\cite{QUA:QUA24521,PhysRevB.67.153106}

At the same time, the astonishing increase in the computational power obtained during the past 40 years has provided a tangible change in the amount of predictions that simulations can provide.

For what concerns \msr, this has also led to the possibility of addressing the three fundamental questions discussed above with fully \emph{ab initio} methods.

In this short review article we will briefly address each question, respectively in Sect.~\ref{sec:where}, \ref{sec:interactions} and \ref{sec:passive}, surveying the importance of the quantum nature of the muon in Sect.~\ref{sec:quantum}. The discussion is illustrated by old and recent examples of DFT aided \msr data analysis. We will limit our attention to the simulations involving positive muons since they constitute the vast majority of the experiments performed with \msr nowadays. Since the topic is still rather vast, particular attention will be devoted to the analysis of the simulations performed in crystalline materials, even though we will also touch a few aspects of the analysis of the muon/sample interactions in molecular compounds.

\section{Where is the muon?}
\label{sec:where}

Part of the first experiments performed with \msr where devoted to the validation of the then new technique.
For this reason the study of elemental crystals and text-book cases were predominant and represented an important development toward a more precise understanding of the muon/sample interactions. Positive muon sites identifications were mainly performed with direct experimental approaches like, for example, following the evolution of the muon frequency shift in a transverse field experiment as a function of the applied stress in a single crystal \cite{PhysRevB.32.293}, notably with the stimulus of theoretical insight \cite{PhysRevB.29.4170}, or by the symmetries of the same shifts as a function of the applied field direction\cite{PhysRevB.30.186}, or again obtaining geometrical constraints on the relative position of the muon and the interacting nuclei from avoided level crossing measurements \cite{PhysRevLett.60.224}.
At the same time, computational methods were used to model the electronic density surrounding the muon and provide, in some cases, information regarding the interstitial position of the muon and the interaction parameters between the muon and its surrounding electronic environment.

\begin{table}
    
\begin{tabular}{ccc}
    \hline
    Compound & \multicolumn{2}{c} {muon site}\\
             & (experiment) & ({\em ab initio}) \\
    \hline
 &&\\
    Fe$_3$O$_4$ & \ \cite{fe3o4,PhysRevB.77.045115} & -- \\
    $R$FeO$_{3}$ ($R$ =  Sm, Eu, Dy, Ho, Y, Er) & \ \cite{PhysRevB.27.5294} & -- \\
    UCoGe & \ \cite{PhysRevLett.102.167003} & -- \\
    YBCO & \ \cite{Brewer_1991,Weber} & -- \\
    UNi$_2$Al$_3$ &  \ \cite{Amato_2000} & -- \\
    GdNi$_5$ & \ \cite{Mulders_2000} & -- \\
    PrNi$_5$  & \ \cite{Feyerherm_1995} & -- \\
    PrIn$_3$ & \ \cite{Tashma_1997} & -- \\
    LiV$_2$O$_4$ &  \ \cite{Koda_2004} & -- \\
    TmNi$_2$B$_2$C &  \ \cite{Gygax_2003} & -- \\
    YMnO$_3$ &  \ \cite{Lancaster_2007} & -- \\
    HoB$_2$C$_2$ &  \ \cite{De_Lorenzi_2003} & -- \\
    UN & \ \cite{M_nch_1993} & -- \\
    UAl$_2$ & \ \cite{Kratzer_1986} & -- \\
    CeAl$_{3}$ & \ \cite{PhysRevB.39.11695} & -- \\
    ZnO &\ \cite{PhysRevB.85.165211} &\  \cite{PhysRevLett.85.1012}\\
    Y$_2$O$_3$ &\ \cite{PhysRevB.85.165211} &\ \cite{PhysRevB.85.165211} \\  
    CoF$_2$, MnF$_2$ & \ \cite{PhysRevB.30.186} & \ \cite{PhysRevB.87.121108} \\
    LiF &  \ \cite{PhysRevB.33.7813} & \ \cite{PhysRevB.87.121108,PhysRevB.87.115148} \\
    YF$_3$ & \ \cite{Noakes1993785} & \ \cite{PhysRevB.87.115148} \\
    MnSi & \ \cite{PhysRevB.89.184425} & \ \cite{jp5125876}\\
    La$_2$CuO$_4$ T & \ \cite{PhysRevLett.59.1045,Hitti1990} & \ \cite {Suter2003329}\\
    La$_2$CuO$_4$ T$^\prime$ & \ \cite{gwenthesis} & \ \cite{pbphd} \\
    $R$CoPO ($R$ = La, Pr)& \ \cite{PhysRevB.87.064401} & \ \cite{PhysRevB.87.064401} \\
    PrB$_2$O$_7$ (B = Sn, Zr, Hf) & \ \cite{PhysRevLett.114.017602} &  \ \cite{PhysRevLett.114.017602} \\
\end{tabular}
\caption{Tentative list of the non elemental crystalline compounds where the muon site is known (or a few candidates are proposed) from the experiment. Although elemental crystal have been largely studied and the muon site is known in most of them, they have been omitted in this table for simplicity and readability.}\label{tab:sites}
\end{table}

In this context, one instructive problem that has been considered with \emph{ab initio} approaches is the diffusion of the muon in copper. This topic has been widely studied both experimentally \cite{PhysRevB.43.3284,PhysRevLett.39.836} and theoretically \cite{PhysRevB.29.5382}.
From the field and orientation dependence of the decay in single crystalline samples\cite{PhysRevLett.39.832} and from \emph{ab initio} calculations it was soon realized that the occupied muon site is the octahedral interstitial.
First principles simulations showed that this is due to the large Zero Point Motion Energy (ZPME) that exceed the self trapping energy gain and makes the tetrahedral interstitial site unstable.\cite{PhysRevB.29.5382} 

Starting from the mid 80s, the increasing computational power allowed to tackle supercells of crystalline materials. Computational efforts were reported mainly for paramagnetic muon states, i.e. muonium, in carbon based and semiconducting materials. Within these simpler crystalline structures \emph{ab initio} methods allowed the identification of the muon sites \cite{0022-3719-17-14-009} and the determination important characteristic of the interaction between the muon and the investigated sample. 

More recently, the advent of large computational facilities 
and the development of effective methods for the solution of the Kohn-Sham equations \cite{PhysRevB.59.1758,PhysRevB.41.7892} 
has led to a rebirth of the DFT approach for providing a description of the muon implantation process in the very end of its deceleration path.
Indeed, since it is rarely possible to determine the muon site(s) with experimental methods, the computational approaches are becoming a precious supporting tool for \msr data analysis.

As of today  the best compromise between accuracy and efficiency to identify muon sites in molecules and crystalline materials is based on a sampling of the total energy hyper-surface which is obtained for the embedding of the charged particle in the host system. 

The starting point is to treat the muon as a hydrogen atom within the standard Born Oppenheimer (BO) approximation used in DFT. 
Depending on the crystalline or molecular nature of the material under study, different approaches are used. In crystalline solids periodic boundary conditions are normally adopted and the final crystal relaxation around the muon is reproduced by choosing a suitably large supercell, in order to limit the interactions between the periodic muon replicas in the simulated system. It is very important to check the convergence of the results against the supercell size. To this aim two requirements are generally inspected:
\begin{enumerate}
    \item the interaction between the muon and its replicas must not influence the results,
    \item the displacement of atoms must progressively decay as the distance from the muon increases. 
\end{enumerate}

This approach has been diffusely used for neutral and charged impurity calculations and has produced accurate results also for \msr  in Si, as well as in other elemental semiconductors \cite{PhysRevLett.58.1547,PhysRevLett.60.2761,PhysRevLett.64.669,Jones351,EPL.7.145}.

For molecular systems the calculation can be less computational expensive since supercells are no longer required. However, in the case of large low symmetry molecules, one must still consider all the possible muon additions to the molecule at inequivalent positions, guiding an educated guess by chemical insight.\cite{doi:10.1021/jp107824p,doi:10.1021/jp305610g}

Table \ref{tab:sites} is a partial list of crystalline compounds where the muon site is assigned, with certainty or tentatively, by experiment. The last rows list the cases where {\em ab initio} confirmed the assignment. 

The stability and the formation energy barrier of the various muon embedding configurations can be estimated, within the same BO approximation, with the nudged elastic-band (NEB) approach. This method provides an efficient way to identify saddle points and minimum energy paths between known initial and final ionic configurations.
Initially, a series of equally separated configurations (called images) along the reaction path is guessed.
The images are kept equally separated from each other along the reaction path by adding an elastic band force acting on the images. A constrained optimization of the total energy for all the images, obtained by projecting out the perpendicular and the parallel components to the path of the spring force and of the true force respectively, provides an iterative method to identify the minimum energy path between the initial and final configurations of the muon and its neighborhood. % in the embedding.

\begin{figure}
\center
    \includegraphics{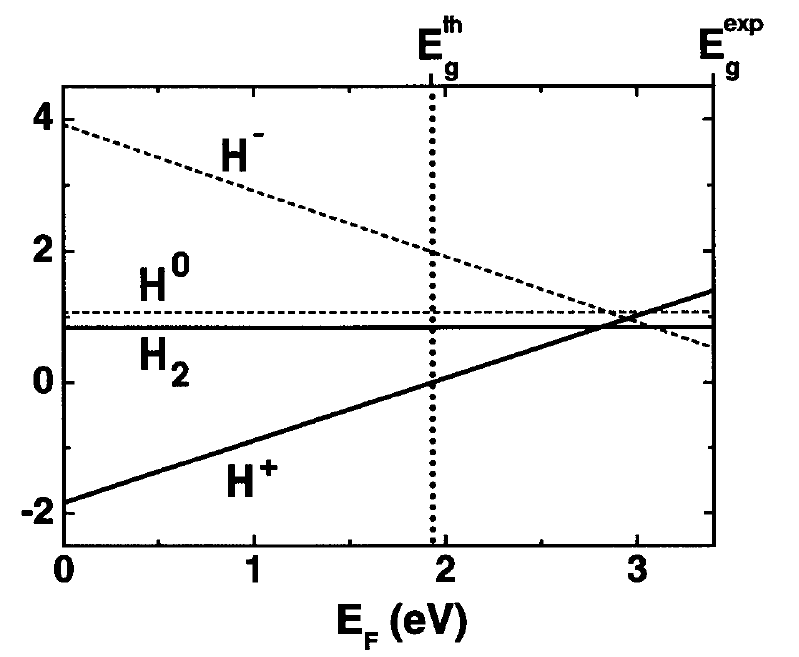}
    \caption{Formation energy for various types of hydrogen impurities as a function of the Fermi Energy level. The plot shows that H$^+$ is the stable form in ZnO, for Fermi Energy values (bottom scales) within the calculated energy gap (top scale), a fact that is experimentally\cite{PhysRevLett.86.2601} verified in ZnO.
    Reprinted figure with permission from Ref.\onlinecite{PhysRevLett.85.1012}. Copyright (2000) by the American Physical Society.}\label{fig:zno}
\end{figure}

One of the first success stories was the discovery of a shallow donor state for hydrogen, hence for the  muon as well, in Zinc Oxide. It was somehow surprising since hydrogen in p-type (n-type) semiconductors typically acts as a compensating impurity, in a stable  H$^{+}$(H$^{-}$) charge state at the bond-center. \cite{PhysRevLett.85.1012}
A 96 atoms supercell affords the calculation of the formation energies of the various charge states of the hydrogen atom and molecule in ZnO. The large H$^{+}$-oxygen bond strength leads to the formation of shallow donor states with a low electron density at the hydrogen (or the muon) site, as it is shown in Fig.~\ref{fig:zno}.  The formation energy depends on three critical parameters: the total energy associated to the impurity in the selected charge state (top axis), the Fermi energy (bottom axis)  and the hydrogen/muon ZPME, discussed in Sect.~\ref{sec:quantum}. The interplay between these quantities governs the stability of paramagnetic and diamagnetic species, although metastable states must also be taken into account in the case of epithermal implanted muons.

The experimental demonstration of the unexpected shallow donor came just one year later with the observation of its characteristic precession frequencies by Cox and co-workers\cite{PhysRevLett.86.2601}, confirming the \emph{ab initio} predictions, and followed two years later by detailed single crystal studies. \cite{PhysRevLett.89.255505}

One common feature of these early examples is that they forcibly concern the simplest crystal structures, that both require more manageable computational efforts, and offer few rather obvious starting guess sites for the muon.
The extension to non elemental compounds with more complex structures also implies a strategy for the exploration of the candidate sites. 

An example of exploration strategy along the same line of ZnO is provided by Vil\~ao  {\em et al.} \cite{PhysRevB.84.045201} who compared  experiments on paratelluride ($\alpha$TeO$_2$) with DFT within GGA on a 3x3x3 supercell (96 atoms) including NEB calculations to discuss diffusion by means of classical energy barriers. They are thus able to identify the experimental results as a donor and a deep trap configuration.

\begin{figure}
\center
    \includegraphics[width=\columnwidth]{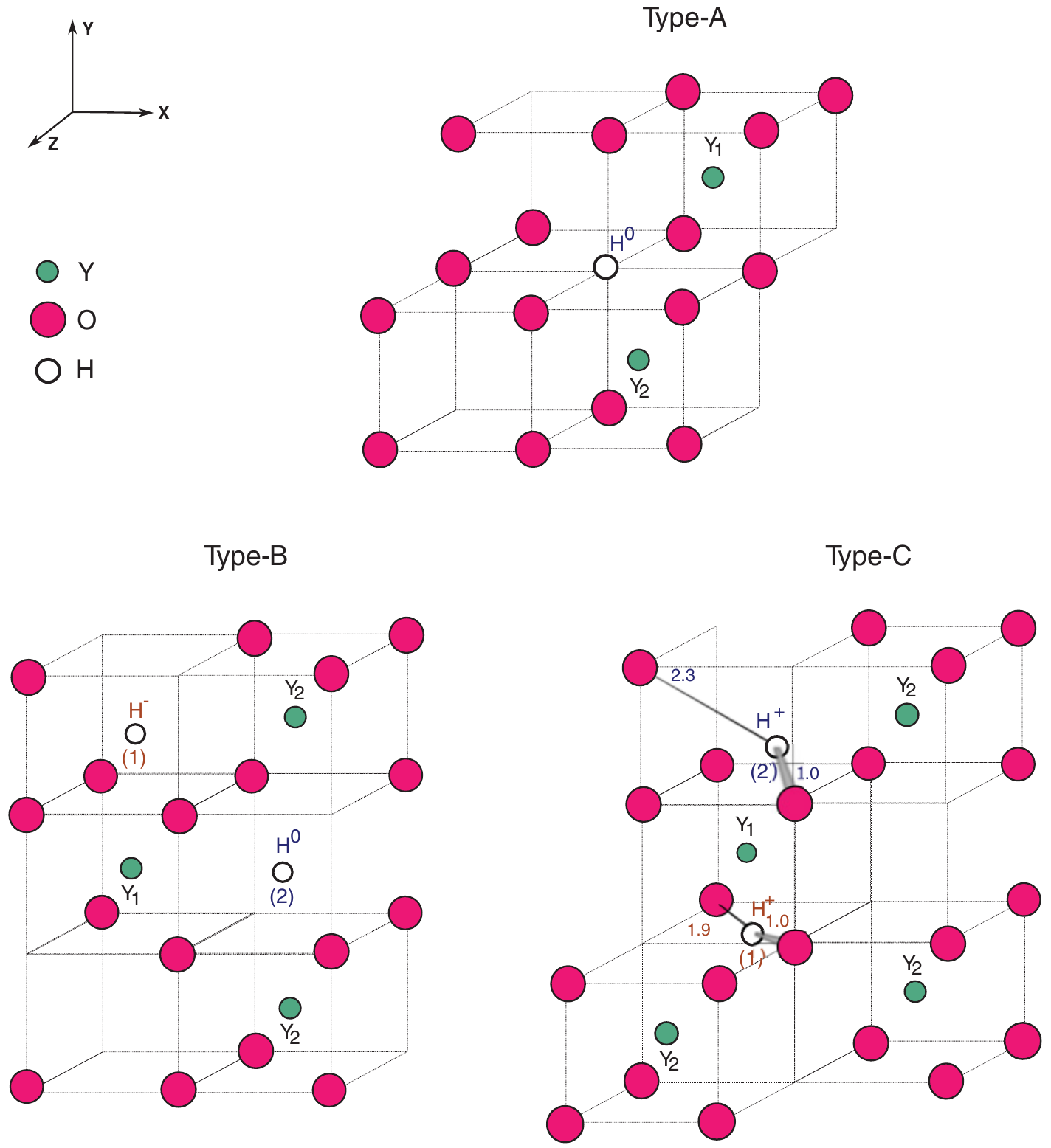}
    \caption{(Color online) Muon sites in Y$_{2}$O$_{3}$. Different positions are identified for the various charge states of the hydrogen impurity used to simulate the muon.
    Reprinted figure with permission from Ref.\onlinecite{PhysRevB.85.165211}. Copyright (2012) by the American Physical Society.}\label{fig:yttria}
\end{figure}

More recently Silva {\em et al.}\cite{PhysRevB.85.165211} assigned the muon species observed in yttria. Figure~\ref{fig:yttria} shows the various diamagnetic and paramagnetic configurations obtained for the different charge states considered for the impurity in the simulations. The results highlight the striking difference between the embedding sites obtained with the different charge states and underline the importance of considering various electronic configurations when exploring the muon embedding sites. These examples are still guided in some degree by the chemical insight allowed by the dominant covalent nature of the bonds. 

Additional strategies are devised for dominantly ionic compounds. Whenever the candidate muon sites are not easily assigned by educated guess it is unavoidable to explore all the possible interstitial sites. To this end it is useful to set up a grid of positions and to optimize the impurity site with the use of the Hellman-Feynman theorem, that provides the intermolecular forces and allows the identification of equilibrium geometries. This procedure produces candidate interstitial positions for the muon.
The number of points in the grid may usually be reduced with the help of symmetry considerations. This can substantially shrink the computational cost of the simulation.

In recent years, DFT based muon site assignment has been used in several materials.
Graphene has been investigated \cite{doi:10.1021/nl202866q} to try and clarify whether diluted hydrogen decorated defects, mimicked by muons, can give rise to magnetism. The authors support their interpretation of the experimental results with DFT predictions of hydrogen decorated carbon vacancies on 6x6 nanoribbon cluster.  By contrast adsorption of positive muon on perfect graphene predicts \cite{Pant_2014}  a ground state energy of 0.22 eV directly on top of the carbon atom. 

Recently, \msr has played a prominent role in characterizing the newly discovered iron based high temperature superconductors. As a consequence, the muon site in these materials has been widely studied.
Most of the initial estimations were based on a naive approach which relies on the analysis of the electrostatic potential of the bulk material obtained from DFT simulations \cite{PhysRevB.80.094524,0953-2048-25-8-084009,PhysRevB.91.144423,PhysRevB.85.064517}.
This method is the direct evolution of the point-charge model widely used in literature.\cite{PhysRevB.84.054430,PhysRevB.85.054111,PhysRevLett.103.147601,PhysRevB.84.064433}
Incidentally we mention that this approach has also been used for materials with diverse electronic properties.\cite{1742-6596-551-1-012052,1742-6596-551-1-012053}
The results obtained from the analysis of the unperturbed electrostatic potential (UEP) are generally validated by comparison with the experiment. 
This is the case of the 11 11 pnictide superconductors\cite{PhysRevB.80.094524}.A full \emph{ab-initio} confirmation of the UEP predictions was provided for the isostructural compound LaCoPO with relaxed supercell calculations.\cite{PhysRevB.87.064401}

The remarkable success of simple electrostatic potential predictions in simpler three dimensional metals may perhaps be justified by heuristic arguments on the screening of point charges. A recent confirmation is the case of MnSi where accurate  supercell calculations\cite{jp5125876} introduce only tiny improvements on the UEP predictions. However the accuracy of the UEP method in a layered material like LaFeAsO or LaCoPO is more surprising, and it should probably be considered just a fortunate case. The method cannot be expected to produce reliable results in insulators or in two dimensional materials alternating metallic and charge reservoir layers. 

Fluorides represent a notable example of insulator where the UEP method fails.\cite{PhysRevB.87.115148} These materials deserve a special mention. Brewer and coworkers identified a striking effect, characteristic of several of them, where the distortion of the lattice produced by the muon on F atoms is very pronounced\cite{PhysRevB.33.7813}. The muon-fluorine distance for the most distorted nearest neighbor ions is characterized in details by the experimental measurement, thanks to the quantum entanglement of the muon and fluorine nuclear spins.

This has been used as an ideal test case by M\"oller {\em et al.} \cite{PhysRevB.87.121108} and by Bernardini {\em et al.} \cite{PhysRevB.87.115148} to verify the accuracy of DFT in reproducing the crystalline distortions introduce by the muon. The first two columns of Table~\ref{tab:Moeller}  show the experimental and calculated muon-fluorine distances, whose very good agreement is an important validation of the \emph{ab initio} approach to the muon site identification problem. 

\begin{table}
    
\begin{tabular}{cccc}
\hline\hline
& $2r_{\rm DFT}$ (\AA) & $2r_{\rm exp}$ (\AA) & ZPE (eV)\\
\hline
(FHF)$^-$ & 2.36 & 2.28 & 0.30\\
(F$\mu$F)$^-$ & 2.36 & & 0.80 \\
LiF & 2.34 & 2.36(2)\cite{PhysRevB.33.7813} & 0.76\\
NaF & 2.35 & 2.38(1)\cite{PhysRevB.33.7813} & 0.76\\
CaF$_2$ & 2.31 & 2.34(2)\cite{PhysRevB.33.7813} & 0.83\\
BaF$_2$ & 2.33& 2.37(2)\cite{PhysRevB.33.7813}& 0.79\\
CoF$_2$ & 2.36 & 2.43(2)& 0.73\\
\hline
\hline    
\end{tabular}
\caption{Calculated (DFT) and experimental (exp) properties of
the diamagnetic and molecular ion  fluorine-muon (F-$\mu$-F) states in solid and vacuum. $r$(\AA) is the muon-fluoride bond length.
Reprinted table with minor edits with permission from Ref.~\onlinecite{PhysRevB.87.121108}. Copyright (2013) by the American Physical Society.}\label{tab:Moeller}
\end{table}

As it has been shown from the very beginning by studying the interstitial muon site in elemental crystals, the ZPME plays a crucial role in this procedure.\cite{teichler1978microscopic,0305-4608-9-7-013,rath1979effect}
Indeed the large ZPME of the muon, discussed in Sect.~\ref{sec:quantum}, may yield classical hopping or quantum tunneling among various interstitial positions. 
The case of fcc copper represent an instructive example in this perspective. \cite{PhysRevB.43.3284} The barrier between the tetrahedral and the octahedral interstitial is too small to bind the muon. Thus no localized muon wave-function is found in that positions. At the same time the barrier between the octahedral sites is small enough to permit both quantum tunneling and classical diffusion for relatively small temperatures.\cite{jp5125876}

In several instances a successful identification of the muon sites based on a first principle approach relies on the additional evaluation of the ZPME, not just on the solution of a purely electronic problem, and the number of cases where this inclusion proved to be essential is steadily increasing.\cite{Suter2003329,PhysRevB.84.045201,PhysRevB.85.165211,Pant_2014,PhysRevLett.107.207207,1402-4896-88-6-068510}

\section{Interaction parameters}
\label{sec:interactions}

Another point that has been largely discussed in the literature is the estimation, from first principles, of the interaction parameters between the muon and the electrons (or possibly the nuclei) of the host material.
Successful results have been reported from early studies of the hyperfine coupling for diamagnetic muon sites in metallic magnetic materials and for paramagnetic muon sites in semiconductors.\cite{PhysRevLett.64.669}
In the vast majority of these preliminary studies the electron density and its magnetic polarization at the muon position were used to estimate the hyperfine parameters according to\cite{PhysRevLett.64.669}

\begin{align}
\label{eq:hyperfine}
    \mathcal{H} & = \sum_i\left\{\frac{2 \mu_{0}}{3}  \mathbf{m}^{\mu} \cdot \mathbf{m}^{e} \delta(\mathbf{r}_i)\right. \nonumber\\ &\left.\quad + \frac{\mu_{0}}{4 \pi} \frac{1}{r_i^{3}} \left[ 3 (\mathbf{m}_{\mu} \cdot \mathbf{\hat{r}_i})(\mathbf{m}_{e} \cdot \mathbf{\hat{r}_i}) - \mathbf{m}_{\mu} \cdot \mathbf{m}_{e} \right]\right\}
\end{align}

where the first term is the Fermi contact term and the second is the dipolar interaction, $\delta$ is the Dirac delta function, $\mathbf{m}^{e} = -g_{e} \mu_{\mathrm{B}} \mathbf{S}^{e}$ and $\mathbf{m}^{\mu} = \gamma_{\mu} \hbar \mathbf{I}^{\mu}$  are the electronic and the muon magnetic moment operators, $\mathbf{r}_i$ is the coordinate of the i-th electron in a reference frame centered at the muon and $\mu_{0}$ is the vacuum permeability.

The parent compound La$_2$CuO$_4$ in its standard crystal group 64 has been first addressed by a 9 Cu + Mu cluster in the Generalized Gradient Approximation (GGA), without quantum treatment of the muon. \cite{Suter2003329} The authors extract spin densities on Cu, nearest neighbor O ions and the muon. They remark that spin density on oxygen, although definitely smaller than that on copper, provides a non negligible contribution through the second term of Eq.~(\ref{eq:hyperfine}), in view of the much shorter distance to the bonded muon, thanks to the $r^{-3}$ dependence of the dipolar term. This remark might be of more general relevance.

For chemically diamagnetic muon species the contact hyperfine coupling at the muon site is generally the result of a rather small imbalance of the spin density, and the accuracy of its determination strongly depends on that of the DFT description of the many body electronic problem. Even assuming small uncertainties of the muon site coordinates and on the electron magnetic moments a large relative numerical error on  the reduced electronic spin polarization at the muon site, produces a considerable relative inaccuracy, that is usually an order of magnitude larger than that typically obtained for the dipolar coupling.

This remark applies e.g.~to the one-dimensional quantum antiferromagnet AF Cu(pyz)(NO$_3$)$_2$, where a 2x2x1 supercell GGA calculation \cite{PhysRevB.91.144417} including the muon in both positive and neutral charge configurations predict large contact couplings for the latter and much smaller ones for the former. The rather low experimental zero field muon precession frequencies rule out the large hyperfine, neutral muon site. The authors implicitly acknowledge the inaccuracy of low spin density DFT estimation by establishing a comparison between the experimental local field and the dipolar contribution alone. This comparison justifies the qualitative statement that the Cu moment must be reduced by a large factor, of order 7-8 with respect to the nominal 1 $\mu_B$, as expected from the low dimensional nature  of the magnetic order. 
\begin{figure}
\center
    \includegraphics[width=0.8\columnwidth]{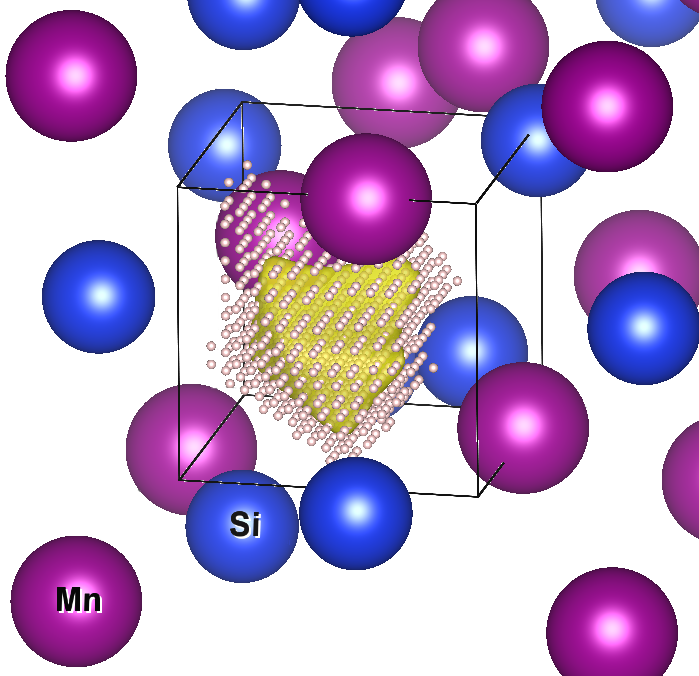}
    \caption{(Color online) The adiabatic muon energy isosurface in MnSi, \cite{PhysRevB.89.184425,jp5125876} obtained joining points where the potential acting on the muon is 1.6 times the corresponding muon ground state energy. The dots represent the points used for the potential interpolation, within DAA (see text).} \label{fig:mnsias}
\end{figure}

Thus, whenever Eq.~(\ref{eq:hyperfine}) provides a reasonably good account of the muon coupling, \msr data together with the muon site assignment allow quantitative determination of magnetic moments and, in fortunate cases, even of magnetic structures. In this perspective a promising approach based on the Bayesian analysis,\cite{Blundell2012113,PhysRevB.89.140413} has been proposed in order to estimate the magnetic moment size and/or the long range magnetic structure of magnetic materials.
Indeed by feeding the probabilistic analysis with the DFT results obtained for the identified muon site(s) it is possible to compare quantitatively the expectations for various spin arrangements and determine the most probable long range magnetic order for the sample under investigation.

Equation (\ref{eq:hyperfine}) represents only a first approximation, since it neglects the quantum nature of the muon. Although the approximation often provides at least the correct order of magnitude, if not the exact experimental value for the hyperfine couplings at the muon site \cite{Katayama1979431,Jepsen1980575}, there are known cases in which this approach is not sufficiently accurate \cite{PhysRevB.25.297, PhysRevB.27.53, PhysRevB.15.1560}.

Finally, we notice that estimating {\em ab initio} the value of the hyperfine field at the muon site may be differently challenging, depending on the material. For strongly correlated electron material it involves an accurate description of their electronic properties that is not always available with conventional DFT approaches. The reliability of the typical approximations, e.g. LDA+U, must therefore be seriously taken in consideration case by case.

At the opposite end rather accurate results can be obtained for the hyperfine coupling parameters of paramagnetic muonated radicals.\cite{ct500027z,PhysRevE.87.012504,C4CP00618F,C4CP04899G,Peck} This is mainly due to the fact that the contact hyperfine coupling constants are proportional, in this case, to the electron density at the muon site, which is much larger for paramagnetic species.
Once again, the key point in this procedure is an accurate electronic description of the molecule, usually obtained with hybrid exchange and correlation functionals.\cite{Peck}

\section{The quantum muon}
\label{sec:quantum}

In the case of both diamagnetic and paramagnetic muon centers, improved results are obtained if the quantum nature of the muon, usually far from being negligible, is taken into account. This was already considered in most of the first reports on \emph{ab initio} studies of the muon \cite{rath1979effect,PhysRevB.25.297}. Since within DFT the muon is treated as a charged classical particle, the simulations must be extended in order to provide a description of the muon wave-function.  

The task is fulfilled with a number of different approaches\cite{1.1356441,Gross1998291,1.3556661,C4CP05192K,PhysRevA.90.042507,PhysRevLett.101.153001,RevModPhys.85.693}. We mention here the estimation of the ground state energy of the muon with the analysis of the phonon-modes, within the BO approximation for the electrons, \cite{PhysRevB.87.121108} and the double adiabatic approximation (DAA)\cite{jp5125876}, discussed in details by Soudackov and Hammes-Schiffer\cite{Soudackov1999503} and Porter {\em et al}\cite {PhysRevB.60.13534} in which an adiabatic energy surface for the solution of the Schr\"oedinger equation of the muon is obtained.

More accurate approaches may be provided by the Nuclear-Electronic Orbitals (NEO)\cite{C4CP06006G}, using Hartree-Fock methods\cite{Kerridge_2004} on an orthogonal basis for the muon and the electrons. Finally, path integral molecular dynamics (PIMD)\cite{1.471221, PhysRevLett.81.1873, Yoshikawa2014135,PhysRevLett.99.205504,PhysRevB.60.14197} in principle can yield the most accurate calculations. However, since these techniques are also listed in order of increasing computational costs, their use grows increasingly impractical for materials, such as mixed valence, strongly correlated electron systems, that are already intrinsically complex from an {\em  ab initio} point of view.

The first two methods treat the muon as a point-like charged particle.
The DAA method and the linear response evaluation of the muon phonon modes in the crystal represent the simplest and less computer intensive approaches. The DAA method approximates the potential of the muon Schr\"odinger equation with the DFT total electronic energy recalculated as a function of fixed muon position in a suitable grid. The phonon based mechanism yields the standard harmonic approximation to the mode frequency, which directly provides the ZPME by a projection method. It has the advantage of treating the muon and the nuclei on the same footing, but the harmonic approximation is usually not very accurate in the muon case. This is indirectly shown for example in Fig.~\ref{fig:mnsias} by the highly non ellipsoidal shape of the muon potential energy isosurface obtained by DAA in the case of MnSi \cite{jp5125876}. The actual shape of the muon potential energy may be mapped within DAA, to avoid the harmonic approximation, but only inasmuch as the energy scales of the nuclei of the embedding system are well separated from those of the muon. Hence DAA may fail for systems with close-lying muon and hydrogen ions. 

The NEO approach introduces a muon wave-function that is optimized together with electronic wave-functions and overcomes the BO approximation, and PIMD treats the nuclei from a quantum perspective by mapping them onto an isomorphic classical polymer of replicas of each nucleus (called bead). In \emph{ab initio} PIMD, each bead requires a self consistent calculation, thus the computational cost scales linearly with number of beads.

\begin{figure}
\center
    \includegraphics[width=0.8\columnwidth]{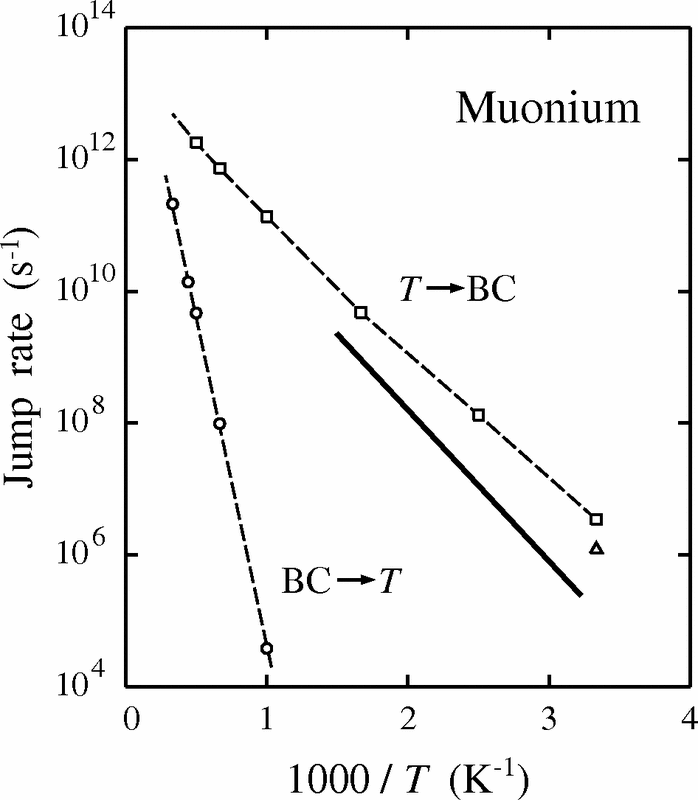}
    \caption{The hopping frequency for a muon in diamond as obtained from PIMD simulations.  The solid line is derived from \msr measurements. Reprinted figure with permission from Ref.~\onlinecite{PhysRevLett.99.205504}. Copyright (2007) by the American Physical Society.} \label{fig:diamond}
\end{figure}

An instance where the drastic approximation of the first two methods (and maybe also of the NEO approach) may fail is the metastability of the neutral H$_0$ charge state in Si, that was first tentatively assigned \cite{Cox1986516} to the tetrahedral (T) site  Mu$^0_T$. Specifically, the existence of an absolute energy minimum at the bond-center (BC) Mu$^0_{BC}$ site \cite{Cox1986516} is confirmed, in Si and diamond, by Hartree-Fock cluster calculations that identify the tetrahedral site as a local minimum.
However, contrasting results regarding the height of the barrier between the two sites and the role of the ground state ZPME of Mu$^0_T$ are reported. \cite{PhysRevLett.71.557, PhysRevLett.58.1547,PhysRevB.36.9122} 

Recalling that thermodynamic equilibrium may not be achieved during a muon lifetime, this finding would agree with the low temperature experimental observation of both species in all elemental semiconductors.
This fact is partially confirmed by more accurate PIMD calculations in diamond.\cite{PhysRevLett.99.205504}
Figure \ref{fig:diamond} displays both the experimental (solid line) and the theoretical PIMD (dashed line) jump rates for muons in diamond. The jump rate from the T site to the BC site is several orders of magnitude larger than that from the BC to the T site for all experimentally accessible temperatures. Therefore the simulation predicts, in qualitative agreement with experiment, that Mu$^0_T$ must disappears from observation at high temperatures, when its jumping rate time becomes shorter than a few nanoseconds, whereas in the same temperature range the more stable Mu$^0_{BC}$ does not delocalise during a few muon lifetimes. 

As of today there is no universal solution to obtain a detailed description of the quantum nature of the muon. While for small molecules PIMD gives the most accurate results, this approach is usually prohibitively time consuming for muons embedded in crystalline materials.
Indeed there are two aspects specific to \msr experiments that makes PIMD very computationally demanding: the small mass of the muon and, in many cases, the (low) temperature that is used in experiments. Both these conditions contribute to the growth of the required number of beads.
For this reason, when performing PIMD, the \emph{ab initio} methods for the electronic structure evaluation are usually abandoned in favor of less demanding approaches like tight-binding Hamiltonians or empirical potentials which may degrade the quality of the description of the electron density.

\section{Is the muon a passive probe?}
\label{sec:passive}

The muon is a positively charged particle\cite{footnote} and, when it comes to rest in the lattice, it gives rise to a charged defect that can introduce appreciable local modification to its immediate electronic and ionic environment. The key question in this case is whether the muon can alter the properties of the system that it is supposed to probe in the experiment.
The answer to this question can be rarely provided just by \msr experiments and it depends largely on the goals of the measurement. Since a large fraction of the muon studies are directed at magnetic materials the question can be often rephrased into ``is the muon distortion capable of altering the apparent magnetic behavior of the investigated sample with respect to that of the crystal without the muon?''

The appreciable crystalline distortions are however generally not a source of concern for magnetic measurements. Their scarce influence may be explained by considering that, firstly, the muon usually forms bonds with the most electronegative atoms of the hosting system, and, in many cases, these atoms do not provide the leading contribution to the exchange integral. Secondly, when the muon does modify the exchange integral, the perturbation is usually confined to the nearest neighbor magnetic atoms. Thus the global magnetic properties are not affected by the perturbation, although the local magnetic field at the muon site may be modified. As long as \msr experiments regard the relative temperature dependence of the local field, the influence is not relevant.

\begin{figure}
\center
    \includegraphics{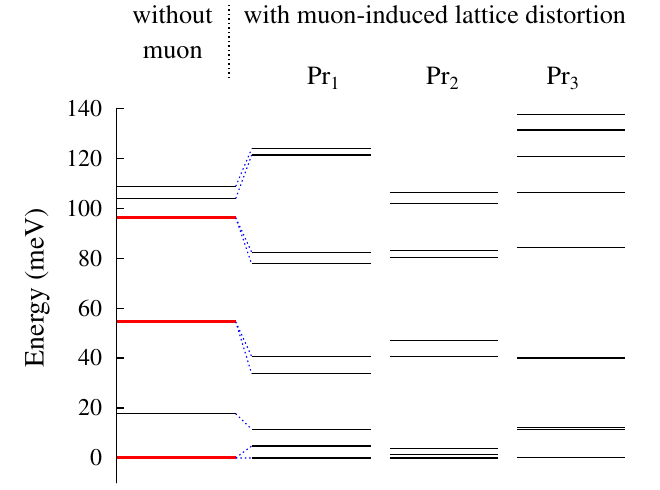}
    \caption{(Color online) Perturbation introduced by the muon on its neighboring Pr atoms in Pr$_2$Ir$_2$O$_7$. The positive electric charge of the muon modifies the crystal field levels producing a non negligible hyperfine coupling at the muon site as a consequence of the ground state doublet splitting. 
    Reprinted figure with permission from Ref.~\onlinecite{PhysRevLett.114.017602}. Copyright (2015) by the American Physical Society.}\label{fig:pr}
\end{figure}

DFT can be directly employed to check the variation of the magnetic moment of equivalent ions as a function of their distance from the muon.  Although most often it is indeed found that the charged particle does not modify significantly the magnetic properties  of its neighbors,
a notably different instance is represented by the one dimensional AF Cu(pyz)(NO$_3$)$_2$ \cite{PhysRevB.91.144417} discussed earlier.
For neutral supercell simulations there are a couple of muon embedding sites which donate an electron to the $3d$ orbitals of their nearest Cu, turning it into the diamagnetic Cu$^{+}$ configuration. Such a substantial local perturbation is, however, still hard to detect.
Indeed, even though the modification does affect the local value of the magnetic field via the dipolar coupling, it has practically no effect on the collective magnetic properties of the system. Therefore the temperature dependence of the magnetic order parameter or that of its slow fluctuations, that induce muon spin relaxation, remain the same as in an unperturbed environment.

By contrast, a very interesting case in which the magnetic response of the system is altered by the muon has been recently discussed by Foronda and co-authors.\cite{PhysRevLett.114.017602} 
An unexpected quasi-static local field at the muon site was observed in geometrically frustrated pyrochlore iridate Pr$_2$Ir$_2$O$_7$. This effect was argued to be related to a muon induced perturbation of the crystal field levels of Pr which leads to the lifting of the non-Kramers degeneracy of the ground state.\cite{1742-6596-225-1-012031}

This hypothesis has been nicely demonstrated by the results obtained with DFT simulations which provides the deformation of the oxygen tetrahedra surrounding the Pr atoms.
The lifting of the degeneracy, shown in Fig. \ref{fig:pr}, is reported for the three Pr atoms closer to the muon. The hyperfine interaction between the Pr nuclei and $f$ orbitals is enhanced by the lifted degeneracy and the resulting magnetic moments of the three Pr atoms surrounding the muon are revealed by the \msr experiment, thus masking the properties of the nonmagnetic unperturbed system.\cite{PhysRevLett.114.017602}

Whenever \msr results deviate drastically from expectations, one should critically consider whether they are due to specific local alteration induced by the muon on its surrounding. It also happens that unconventional muon related phenomena are invoked to justify \msr observations. These cases  too may profit from a comparison with DFT predictions. For example, it has been suggested that spin polarons may interact strongly with muons, in particular in the noncentrosymmetric magnetic metal MnSi \cite{PhysRevB.83.140404}. In this material the spin polaron would consist of an electron coupled to four Mn neighbors, to form a single large spin entity. Binding of the muon to the spin polaron was claimed to justify two precessions observed by Storchack {\em et al.} in high transverse magnetic field. \cite{PhysRevB.83.140404} This explanation is alternative to the conventional hypothesis of a muon site made inequivalent by the application of the field, to justify the appearance of more than one frequency.  For MnSi the site identification by DFT, supporting accurate transverse field experiments and a careful data analysis, proved that the observed frequencies correspond to the latter case. \cite{PhysRevB.89.184425,jp5125876}

A similar instance is that of deconfined magnetic monopoles that are predicted by theory in spin ices, such as in some rare earth pyrochlores. A recent experiment claimed to have detected by \msr in Dy$_2$Ti$_2$O$_7$ a second Wien effect, also referred to as magnetricity, i.e.~the dissociation of magnetic charges by an applied field. \cite{nature461.956} A subsequent work \cite{PhysRevLett.107.207207} demonstrated this not to be the case, by comparing observed and simulated spectra on the same material, based on DFT site assignment. Incidentally this is a case where the error in interpretation of the earlier experiment turned out to be a trivial one, and its recognition did not actually rule out the existence of deconfined magnetic charges in Dy$_2$Ti$_2$O$_7$. Rather, a more recent work suggested \cite{PhysRevLett.108.147601} that the observations of Bramwell {\em et al.} \cite{nature461.956} are probably still due to magnetricity, although the observation in that work was rather indirect, through a large fraction of muons implanted in close contact to the sample, but outside it, in its cryostat holder.

In both the MnSi and the Dy$_2$Ti$_2$O$_7$ cases qualitative analysis could suffice to produce plausibility arguments towards the correct conclusion, but the DFT offered a precious quantitative support to the discussion of the experimental findings. This and other examples,\cite{PhysRevB.87.121108} show the effectiveness of the computational approaches in confirming or rejecting the generally accepted belief that the muon behaves as a passive probe.

\section{Limits and Perspectives}
One can confidently say that DFT calculations nowadays offer methods for predicting muon candidate sites in many crystalline materials, suitable to be directly employed in the design of \msr experiments and to provide complementary information for the data analysis. We refer here to muon {\em candidate} sites because it is presently beyond the scope of DFT to actually predict the branching ratios among such sites during muon implantation, which is effectively an epithermal process.   

The literature on DFT based analysis of the effect of charged impurity is vast and often provides precious guidance for the validation of the muon results obtained by numerical simulations. Three intrinsic limiting factors arise when considering DFT as a tool for complementing muon experimental observations.

Open challenges are still represented by critical compositions of solid solutions, such as certain intermetallics, or intermediate valence oxides. It is for instance still very hard to accurately simulate by DFT a specific composition like YBa$_2$Cu$_3$O$_{6.35}$, at the onset of high T$_c$ superconductivity, since a very large supercell would be required. But enough insight can be often gathered by considering end members and simple intermediate compositions.   

Another difficulty is sometimes caused by the mean field approach of the Kohn-Sham method, not always sufficient to describe the electronic properties of the materials in all their relevant details. This is notoriously true already for semiconductors, e.g. when excited states are involved, as for the energy gap, although in this case the shortcomings  for the muon are easily circumvented. Failures may be more difficult to overcome in the case of strongly correlated systems. 

A third problem is related to the BO approximation that is commonly adopted when simulating the muon with \emph{ab initio} approaches. This latter issue may become rather severe when dealing with the interaction between light atoms and the muon.

Since the purpose of the present review is to concentrate on methods that may be routinely available to assist the analysis of \msr experiments, not all the known theoretical tools qualify. In this sense a universal viable way to tackle the quantum nature of the muon is still missing. Although from the theoretical point of view, many approaches have been developed,\cite{gwfornonbo,1742-6596-225-1-012031,1.471221}  most of them become computationally increasingly expensive with the number of atoms, and the number of electrons per atom that are included in the calculation. A compromise between accuracy and speed must be found. 

Promising results have been obtained with the nuclear-electronic orbital (NEO) method \cite{jp053552i,C4CP06006G} which, by treating only a small subset of the atoms with non-BO approaches, improves the description of the muon and of other light nuclei with smaller computational costs with respect to other approaches like, for example, PIMD.
For the cases where the computational cost of performing PIMD is sustainable, this approach has demonstrated high accuracy\cite{ct500027z}.
Nonetheless, as of today, its applicability is limited to simpler systems such as molecular materials, where the single DFT self consistent field simulation is computationally not expensive. 

Whenever the muon coupling to its environment is dominated by dipolar interactions the site assignment is already sufficient to obtain a fully quantitative muon data analysis, and the influence of the quantum muon treatment may be much less important. By contrast, when contact hyperfine couplings are required for a chemically diamagnetic site, state of the art DFT techniques may not be sufficiently accurate, although the situation will probably change during the next years owing to the advances of both computational methods' efficiency and computational power availability.

Finally, we have shown that the methods we have described above are already a very valuable tool when critically analyzing the possibility of a muon induced effect. 

\begin{acknowledgment}

The seed for both our recent DFT work and the effort to collect the systematics which is the basis of the present review was generated by the MUON JRA of EU FP7 NMI3, under grant agreement 226507. We would like to thank Franz Lang, Johannes M\"oller, Stephen Blundell and Fabio Bernardini for useful discussions. We also acknowledge partial funding from PRIN project 2012X3YFZ2 and from the European Union’s Horizon 2020 research and innovation programme under grant agreement No 654000. 

\end{acknowledgment}

\bibliographystyle{jpsj}

\begin{thebibliography}{100}
\bibitem{footnote}
Note that the positive muon, that is a lepton according to the Standard Model,
  is an anti-particle, whereas the negative muon is the particle. This has
  virtually no consequences on the experiments, since the chance of
  particle-anti-particle annihilation is totally negligible for muons. Instead,
  the negative muon fate in atomic matter tends to be strongly influenced by
  nuclear capture, with a large impact on its average lifetime. Since the muon
  decay cannot be accurately accounted a-priori in this case, it is then much
  harder to extract the spin relaxation function for negative muons. Negative
  muon spin spectroscopy tends therefore to be a niche application.
\end{thebibliography}

\end{document}